# Carrier-envelope phase stabilization with a petahertz optical chip

*Václav Hanus[1], Michał Pietrzak[2], Jarosław Sotor[2], Péter Sándor[1], Ferenc Krausz[3,4,5,6], Péter Dombi[1,7,*], Maciej Kowalczyk[2,#]*

[1]HUN-REN Wigner Research Centre for Physics, 1121 Budapest, Hungary
[2]Faculty of Electronics, Photonics and Microsystems, Wrocław University of Science and Technology, 50-370 Wrocław, Poland
[3]Ludwig-Maximilians-Universität München, 80539 München, Germany
[4]Max-Planck-Institut für Quantenoptik, 85748 Garching, Germany
[5]Center for Molecular Fingerprinting, 1093 Budapest, Hungary
[6]The University of Hong Kong, Chong Yuet Ming Physics Building, Pokfulam Road, Hong Kong
[7]Department of Atomic Physics, Budapest University of Technology and Economics, 1111 Budapest, Hungary

*E-mail: dombi.peter@wigner.hu

#E-mail: maciej.kowalczyk@pwr.edu.pl

## Abstract

Stabilization of the carrier-envelope phase (CEP) in trains of laser pulses is essential for generating optical frequency combs for ultraprecise metrology and for studying ultrafast, lightwave-controlled laser-matter interactions. Conventional CEP detection is based on nonlinear interferometry, which infers the phase indirectly from a beat-note averaged over many pulses, consumes the sampled light, and requires bulky, wavelength-specific optics preventing integration. In contrast, lightwave-driven currents provide direct and single-shot conversion of the optical waveform to electrical signal, but it has been so far employed only for passive CEP detection and mapping. Here, we extend the use of lightwave-driven currents beyond passive CEP sensing to active CEP control. A solid-state, micrometer-sized petahertz (PHz) optical chip converts the CEP of the incident pulses into a feedback voltage signal, which is employed for interferometry-free waveform stabilization. We induced lightwave-driven currents by mid-infrared pulses for the first time, at electric-field amplitudes as low as 0.01 V/Å, and we successfully used this approach to CEP-stabilize a 2.3-µm Cr:ZnS oscillator delivering nanojoule pulses at 25 MHz repetition rate. We measured a residual phase jitter of 150 mrad in the laser output. The simple and transmissive chip-scale architecture is not restricted to a specific spectral range and leaves the full beam available for experiments. Our work establishes a direct electrical route towards integrated sources of waveform-stable few-cycle pulses for ultrafast science and PHz electronics.

## Introduction

The generation of infrared light pulses possessing only a few oscillation cycles of the electric field, together with sub-cycle control over their optical waveforms[1–4], has opened access to attosecond physics[5–7], probing electron dynamics on its natural timescales[8–10], and to extreme nonlinear optics allowing strong-field control of light-matter interactions[11,12], including nano-optics[13,14]. These interactions constitute the physical basis for lightwave electronics, where controlled optical fields directly drive charge motion[15,16], and thereby offer a route towards petahertz (PHz) signal processing far beyond conventional electronic bandwidths[17,18]. The enabling technology is the stabilization of the carrier-envelope phase (CEP) ensuring pulse-to-pulse reproducible optical waveforms. Complementarily, in the frequency domain, CEP stabilization is equivalent to fixing the offset of the optical frequency combs, which has advanced optical frequency metrology to unprecedented levels of precision[19–21].

CEP control requires the detection of the carrier-envelope offset frequency ($f_{ceo}$), which, since the pioneering demonstration of optical waveform control[2–4], relies on nonlinear interferometric techniques such as *f*-to-2*f* self-referencing[1]. These approaches require octave-spanning spectra generated through nonlinear broadening and frequency conversion stages, increasing experimental complexity and instability. *f*-to-2*f* schemes also limit scalability and compatibility with integrated architectures. The construction of a *f*-to-2*f* interferometer requires critical dichroic components fitted for the wavelength of the laser such as mirrors, beamsplitters, and nonlinear crystals that can only be used with a certain laser type. Moreover, optical power sent to the interferometer is unavailable for downstream experiments. Finally, because spectral interferometry infers CEP indirectly, it does not provide direct electronic access to the optical waveform on a single-shot basis at MHz rates. Beyond spectral interferometry, optical waveforms can be measured by their encoding into sub-cycle electron dynamics in techniques relying on strong-field interactions such as attosecond streaking[22,23], or above-threshold ionization (ATI)[24,25]. These typically rely on gas ionization by kHz-repetition-rate, high-energy (multi-μJ) pulses in a vacuum chamber and require electron spectroscopy, which limits their applicability as compact CEP sensors for the broader class of moderate-energy laser sources.

A promising route towards overcoming these limitations is provided by lightwave-driven currents in solids[15,16,26]. This phenomenon enables the electric field of a few-cycle laser pulse to induce charge motion on sub-cycle timescales, allowing to convert the driving optical waveform directly into an electrical signal. Since changing the CEP modifies the temporal symmetry of the electric field, CEP-dependent currents provide a direct electrical signature of the optical waveform. This principle was demonstrated in the seminal work of Schiffrin et al., who showed CEP control of lightwave-driven currents in fused silica[15]. Subsequently, experiments by other researchers showed relaxed demands on material choice and laser parameters. The currents have been observed upon an interaction with various material types including dielectrics[15,27,28], semiconductors[29–32], metals[33] and 2D materials[17,34,35]. At the same time few-cycle, non-amplified compact oscillators with pJ-level pulses were shown to be sufficient to drive CEP dependent currents[32,33]. Comparable downscaling of CEP sensitivity has otherwise been achieved only via plasmonic enhancement in nanoantenna junctions[14,36,37]. To the best of our knowledge, all prior demonstrations (not relying on the plasmonic enhancement) have been so far realized by near-infrared drivers around 800 nm[18]. Extending lightwave-driven current generation to longer wavelengths should provide higher yield as the quasi-static assumption for the electron acceleration is better fulfilled[38].

Together, these aspects highlight the necessity of a simple, wavelength-independent optical chip for CEP stabilization requiring only two ingredients: a sufficiently strong few-cycle optical field driving waveform-dependent charge motion in a solid-state medium and nearby metallic contacts collecting the resulting current. With sufficient signal-to-noise ratio and electronic bandwidth, compact direct electrical readout of the optical waveform can be realized that is naturally compatible with feedback control for CEP stabilization. Despite rapid advances in lightwave PHz electronics in recent years[18], CEP-sensitive currents have so far served only as passive waveform sensors enabling pulse-to-pulse phase tagging[26], spatial CEP mapping[39], or field samplers based on nonlinear photoconductive sampling (NPS)[40]. Harnessing PHz currents as a direct feedback signal for CEP stabilization has remained an open challenge.

In this work, we demonstrate active CEP stabilization using lightwave-driven currents generated in a solid-state, micrometer-scale PHz optical chip. The chip provides direct CEP-to-voltage conversion and delivers the feedback signal, without using conventional interferometric detection. We bring lightwave-driven currents to the mid-infrared (mid-IR) range for the first time, and use them to stabilize the CEP of a 2.3-μm Cr:ZnS oscillator output delivering nanojoule pulses at MHz repetition rates, reducing the residual phase jitter to 150 mrad. Mid-IR excitation enables current generation at electric-field amplitudes as low as 0.01 V/Å[32], extending lightwave-based CEP detection to weak driving fields. Because the ultrafast currents are measured in a transmissive geometry, the driving light remains available for downstream experiments. Our results extend lightwave electronics from passive sensing to active CEP control, establishing a compact, directly electrical route to integrated sources of waveform-stable few-cycle pulses for frequency-comb spectroscopy[20] and strong-field science[6,38,41].

## Results

*Experimental setup*

Our experiment combines an ultrafast mid-IR Cr:ZnS oscillator[42,43] with a light-field-driven PHz chip[32,33] for direct measurement and stabilization of the laser's CEP. The experimental setup is schematically presented in Fig. 1(a). When mode-locked, our laser generates 26 nJ pulses at a central wavelength of 2.3 μm at a repetition rate of 25 MHz. Subsequently, the oscillator output is focused on a $TiO_2$ plate for spectral broadening to cover two-octave-spanning bandwidth with 18 nJ, see Fig. 1(b). A fraction of the full power is split and sent to an *f*-to-2*f* interferometer for traditional out-of-loop $f_{ceo}$ beat-note detection[42]. By a combination of custom-designed chirped mirrors and material dispersion[44,45] the pulses are compressed down to a two-cycle duration of 14 fs, see Fig. 1(c), as determined by a FROG measurement (see Supplementary Information). A pair of $BaF_2$ optical wedges implemented at the output of the compressor provides precise control of the pulse duration to ensure the highest intensity on the sample and to tune the CEP. The beam containing pulses with up to 10 nJ energy is focused by a parabolic mirror to a 7-μm spot on the PHz chip.

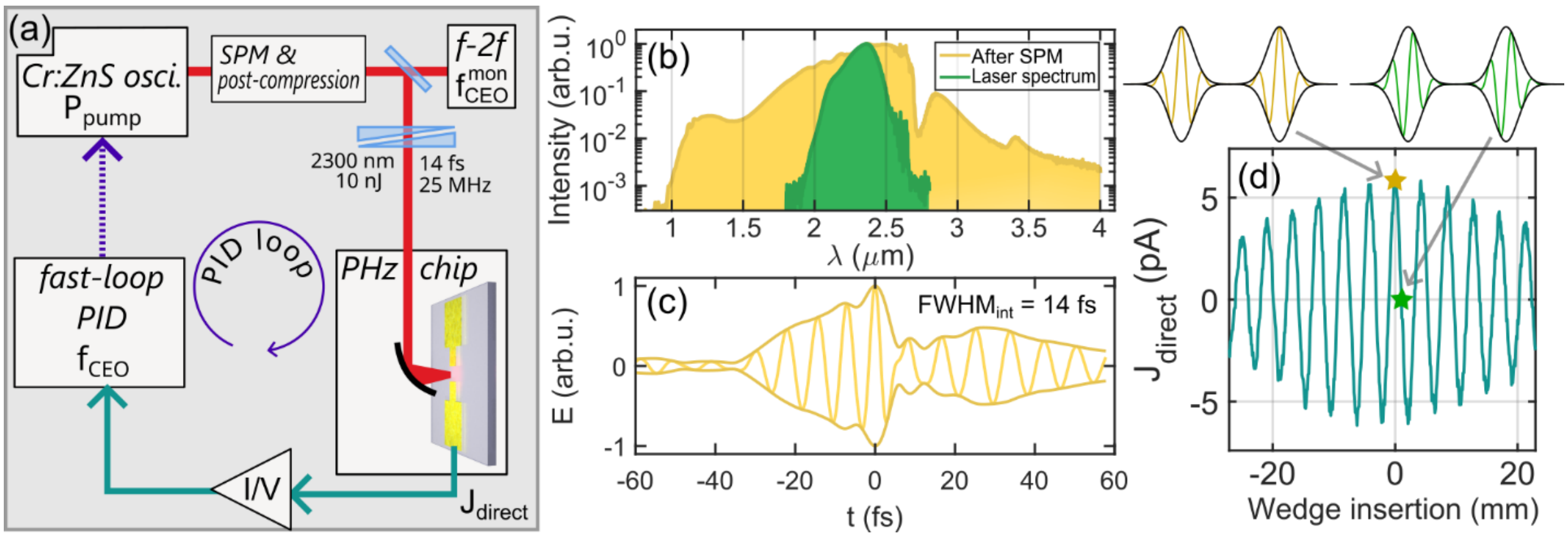

**Fig. 1. | Experimental setup and CEP-to-current conversion.** ***(a)*** *Scheme of the experimental setup. The Cr:ZnS laser oscillator output (25 MHz, 2.3 μm) is spectrally broadened and post-compressed, then split into two arms: one is lead to a conventional f-2f interferometer; in the other arm the pulses with 10 nJ energy are compressed using* $BaF_2$ *wedges to 14 fs and focused onto a PHz chip consisting of a micrometer-sized gap between two gold electrodes on an active medium which is either bulk GaN or an Ir-based nanolaminate. The few-cycle pulses drive CEP-dependent current across the gap, which is converted into voltage providing the error signal for a proportional-integral-derivative (PID) servo loop that phase-locks the oscillator.* ***(b)*** *Optical spectrum of the Cr:ZnS laser before (green) and after (yellow) spectral broadening.* ***(c)*** *Temporal profile of the electric field E of the compressed 14-fs pulses as retrieved from a FROG measurement. The absolute phase of the depicted carrier is set to zero for illustration purposes.* ***(d)*** *Photocurrent* $J_{direct}$ *generated in the PHz chip versus* $BaF_2$ *wedge insertion, which modulates the driving-field CEP.*

*Lightwave-driven currents in the PHz chip*

Focusing conditions together with the sufficient peak power provide on-target amplitude of the electric field strength of about 0.2 V/Å that is sufficient to generate lightwave-driven currents. The laser beam is positioned with its focus on the active material of the PHz chip while passing through a gap between two metallic electrodes. The gap size is between 1 and 5 μm depending on the sample considered, see Methods. The laser-matter interaction gives rise to the CEP-dependent current across the gap, which is detected using a transimpedance amplifier. Notably, the samples are transmissive, hence the light driving the ultrafast currents can be readily used for downstream experiments.

As a first step, we CEP-stabilized the laser via conventional *f*-to-2*f* interferometry. By setting $f_{CEO} = 0$, the current signal from the PHz chip, $J_{direct}$, produced a constant DC readout voltage that varies with CEP. This is illustrated in Fig.1(d), which depicts variation of the current generated in the chip, due to gradual insertion of the upstream wedge causing periodic CEP change. Moreover, excess material dispersion of the wedge introduces chirp to the driving few-cycle pulses, leading to a gradually decreasing oscillation amplitude[30], as already observed in early photoemission experiments[46,47]. If $f_{CEO}$ is set to a non-zero value, the resulting pulse-to-pulse CEP evolution can be directly traced on an oscilloscope (Fig. 2(a), $f_{CEO}$ = 40 kHz).

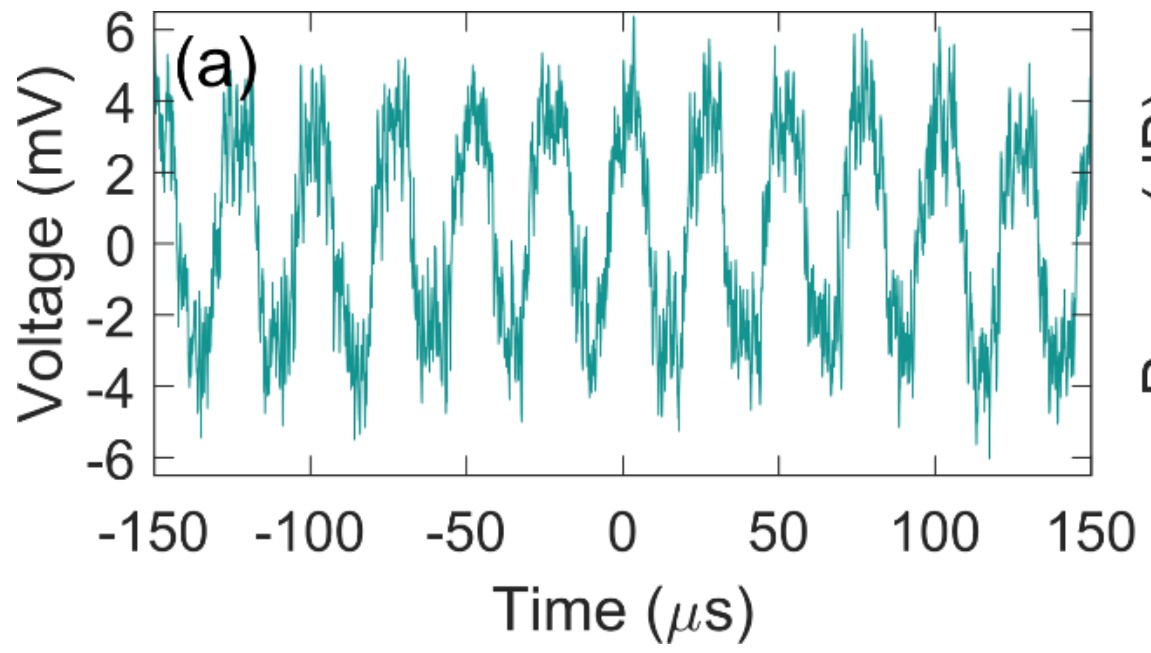

**Fig. 2. | Electrical readout of lighwave-driven currents from the PHz chip.** ***(a)*** *Pulse-to-pulse CEP evolution recorded by an oscilloscope with* $f_{CEO}$ *set to 40 kHz.* ***(b)*** *Free-running* $f_{CEO}$ *beat note recorded by a radiofrequency (RF) spectrum analyzer.*

*Direct CEP stabilization via lightwave-driven currents in the PHz chip*

To evaluate lightwave-current-based CEP stabilization, we disengaged the conventional interferometry-based $f_{CEO}$ stabilization and monitored the free-running CEP change with an RF spectrum analyzer (Fig. 2(b)). Although the CEP-induced voltage oscillations from the PHz chip can be tracked in a single-shot manner, here we were limited to $f_{CEO}$ frequencies up to ~1 MHz, due to bandwidth limitations of the employed transimpedance amplifier (see Methods). As the response of the chip to laser light is intrinsically sub-cycle, faster readout electronics would support single-shot CEP detection.

In order to close the CEP-feedback loop, the current signal from the PHz chip was amplified and mixed with an RF reference from a signal generator. The generated error signal is fed to a proportional–integral–derivative (PID) servo controller enabling CEP stabilization via pump-power modulation (see Methods). Once the CEP lock was established by engaging the feedback loop, stable operation was maintained continuously during extended uninterrupted operation.

The performance of the CEP lock, with $f_{ceo}$ locked to 200 kHz, is shown in Fig. 3 presenting the RF spectrum and the corresponding phase-noise power spectral density. The residual phase noise is measured out-of-loop in the *f*-to-2*f* interferometer with an integrated value of 150 mrad (in the frequency range from 10 Hz to $10^5$ Hz ($f_{ceo}/2$)), see Fig. 3(b). We emphasize that the CEP lock was achieved exclusively using the signal from the PHz chip, without any assistance from the interferometer, which was employed solely for out-of-loop phase noise characterization.

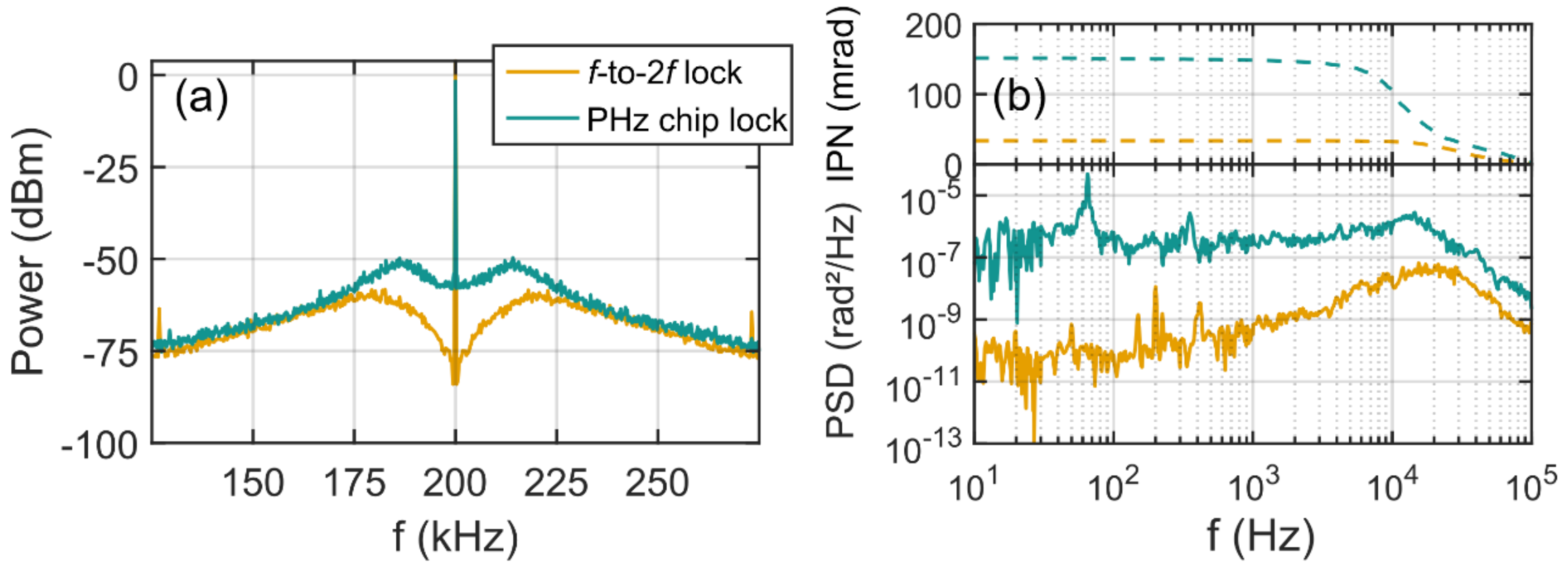

**Fig. 3. | Active CEP stabilization using lightwave-driven currents from the PHz chip.** ***(a)*** *Out-of-loop RF spectrum of the* $f_{CEO}$ *beat-note locked to 200 kHz via lightwave-driven currents (green) and, for comparison, conventional f-to-2f interferometry (yellow).* ***(b)*** *Corresponding phase noise power spectral densities (PSD, lower panel) and integrated phase noise (IPN, upper panel) with same color coding. All measurements were performed out-of-loop with an independent f–to-2f interferometer.*

The residual CEP noise is currently limited by the strength of the lightwave-driven currents generated in the PHz chip, which determines the signal-to-noise ratio SNR of the detected $f_{ceo}$ beat-note (Fig. 2(b)) and hence the achievable lock performance[48]. To verify this interpretation, we studied the performance of the CEP lock based on conventional *f*-to-2*f* interferometry (yellow lines in Fig. 3). The resulting residual-noise spectrum exhibited a similar profile, with the servo-bump cut-off shifted by only 5 kHz (to 20 kHz), while achieving noise suppression down to 35 mrad. These results indicate that further scaling of the lightwave-driven currents generated in the PHz chip would directly improve the SNR of the detected $f_{ceo}$ beat-note, and enable CEP-lock performance comparable to the conventional interferometric technique.

*CEP-dependent currents in the mid-IR regime*

Here, we drive the CEP-dependent currents with two-cycle pulses at 2.3 µm central wavelength. To the best of our knowledge, all previous demonstrations of the light-field-driven currents in bulk media have relied on near-IR sources, typically Ti:sapphire systems providing few-cycle pulses around 0.8 µm[18]. Here, we would like to highlight the advantage of driving at 2.3 µm resulting in substantially higher current yields. To verify the advantage, we calibrated the on-target amplitude of the electric field of light and measured the yield of the current extracted from the PHz chip (see Methods). Subsequently, we examined a range of field strengths spanning from 0.01 V/Å to 0.25 V/Å, as shown in Fig. 4. The signal appearance field-strength threshold for the GaN-based PHz chip driven by a two-cycle mid-IR pulse is as low as 0.01 V/Å, see Fig.4(c). This field strength is 20 times lower than the previously reported minimum of 0.2 V/Å, achieved at a central wavelength of 0.8 µm by Hanus et al.[32]. To the best of our knowledge, it is the lowest field strength at which lightwave-driven currents have been observed under direct excitation, excluding approaches exploiting electron tunnelling over nanojunctions[49].

To test whether the benefits of mid-IR driving extend beyond a single material platform, we present the field-strength scaling in two representative material classes. We investigated semi-insulating GaN and a metallic nanolaminate heterostructure containing Ir metallic and $Al_2O_3$ layers (nano-Ir[33]). Experimental scaling data are accompanied by power-law fits. There is an apparent difference in field amplitude scaling of the current $J_0(E_0)$: The power-law fit $E_0^q$ applied on the data shows different exponents $q$ for the two examined materials as it results in values of 1.9 (Fig.4(a)) and 3.3 (Fig.4(b)) for GaN and nano-Ir respectively. The difference in the scaling can be attributed to a different mechanism underlying the CEP-dependent current generation. In GaN the interband multiphoton valence-to-conduction band coupling and activation of subband-gap virtual states can play the leading role[50]. The presence of metallic layers in the nano-Ir sample hints rather on intraband mechanisms that lead to the CEP-dependent current originating from non-quadratic band potential and Bloch oscillations[33]. Resolving the underlying microscopic phenomena is beyond the scope of this work and will be the focus of our future studies using complementary experimental methods with sub-cycle temporal resolution.

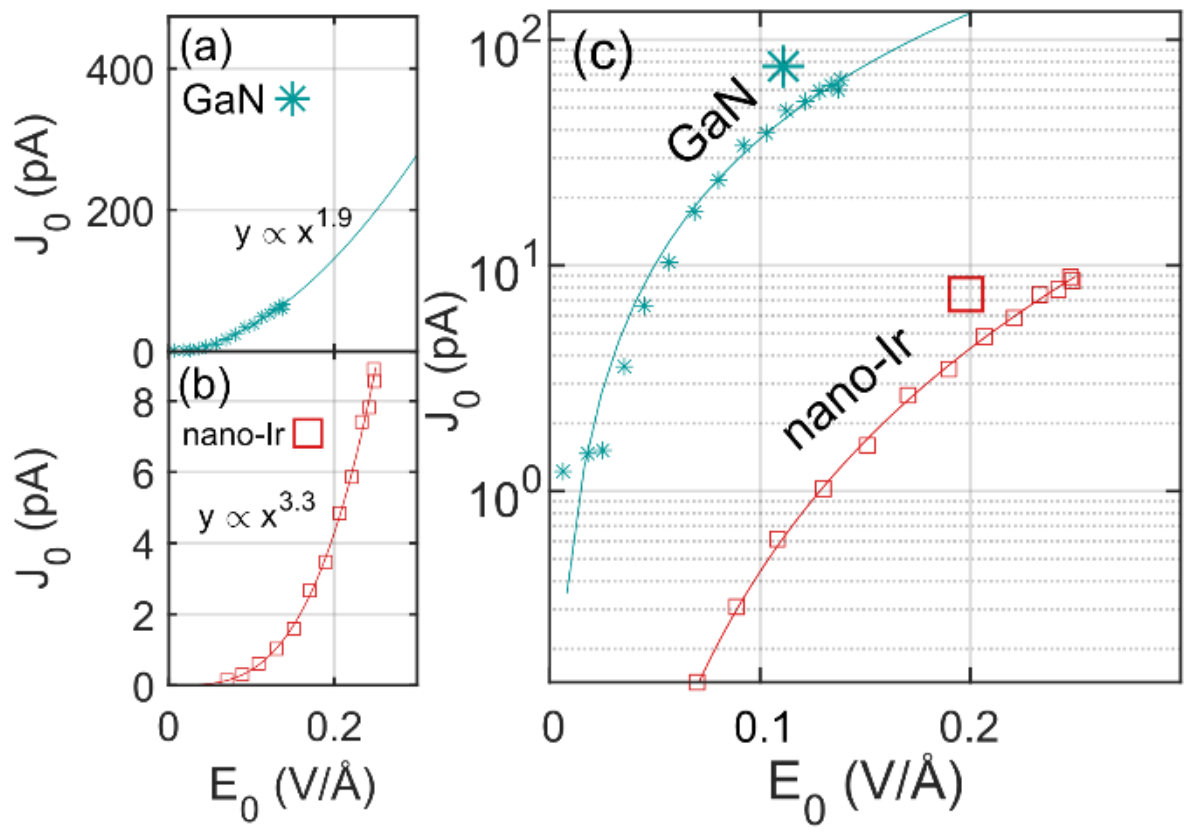

***Fig. 4. | Electric field-strength scaling.*** *Field amplitude $E_0$ scaling of the CEP dependent current from the PHz chip detected as an amplitude $J_0$ from a lock-in amplifier. The measured data are presented in a linear scaling for GaN (a) and nano-Ir (b) samples and they are fitted with power-law function. (c) Comparison of current yields from (a,b) in a logarithmic plot accompanied by the power-law fits.*

## Discussion

We have demonstrated active CEP stabilization using lightwave-driven currents generated in a solid-state PHz optical chip. By enabling direct CEP-to-voltage conversion, the chip provided the feedback signal required to close the servo loop without conventional interferometric CEP detection. As a substantial advantage, our approach is not intrinsically tied to a specific wavelength range and the same chip-scale architecture can be applied for different laser systems. Because the underlying charge dynamics are sub-cycle, the approach is intrinsically single-shot; limited only by the bandwidth of the readout electronics. Finally, unlike interferometric self-referencing, using transmissive PHz chips for CEP detection is nondestructive and enables using the full laser beam for downstream experiments.

Here, we implemented this concept by generating, to our knowledge, the first lightwave-driven current control scheme in the mid-IR spectral domain and used it to actively CEP-stabilize a 2.3-µm Cr:ZnS laser oscillator delivering nJ-level, few-cycle pulses at 25 MHz repetition rate. We prove that using PHz chips for CEP stabilization is not constrained to a single material class, by realizing it in semiconducting GaN and metal (Ir) nanolaminate platforms. The residual CEP noise of the stabilized Cr:ZnS laser (150 mrad) is currently limited by the SNR of the detected $f_{ceo}$ beatnotes, rather than by any intrinsic limitation of our concept.

Further scaling of the CEP-dependent current signal has the potential to improve the CEP lock towards the current state-of-the-art performance of the conventional interferometric approach[43] and several routes are available to achieve this goal. Optimizing the PHz chip material, electrode geometry and gap size can directly increase the collected charge per pulse. In addition, we observed optically induced currents at an electric-field strength 20 times lower than the previous record threshold achieved with near-IR Ti:sapphire lasers[32]. This indicates that the signal level can be increased by extending the driving wavelength, making the few-cycle mid-IR drivers crucial for scaling lightwave-controlled PHz electronics from proof-of-principle demonstrations to practical implementation[18].

A further important potential, yet to be exploited, is that the reduced electric field strength needed to generate the phase-sensitive current paves the way towards extending CEP control to much higher laser pulse repetition rates, possibly approaching or even exceeding the GHz frontier for this parameter. This can establish new bitrate regimes in PHz electronics platforms.

In summary, our work transforms the lightwave-driven optical chip from a passive waveform sensor into an active control element, establishing a direct optical-electrical solution for CEP stabilization. This opens a path towards integrated sources[51,52] of frequency combs for ultraprecise metrology[20] and waveform-stable few-cycle laser pulses for strong-field science and petahertz electronics[53].

## Methods

### Cr:ZnS laser system

The laser system is a soft-aperture Kerr-lens mode-locked laser oscillator based on a 5-mm-long Cr:ZnS gain crystal pumped by a custom 1.56 µm fiber amplifier optimized for minimal amplitude noise[43]. The output of an additional low-power (100 mW) pump diode is combined with the main pump beam using a polarization beam combiner, enabling CEP stabilization via pump-power modulation. A pair of intracavity wedges enables slow CEP correction, when the available range of the modulation diode is reached. The output of the Cr:ZnS laser (650 mW) is focused to a ~10 µm spot in a 0.5-mm-thick, [001]-cut, AR-coated $TiO_2$ plate. Spectral broadening is accompanied by filament formation, introducing propagation losses[42] and reducing the output power to 450 mW. Subsequently, the beam is incident on a ZnS wedge acting as beamsplitter based on Fresnel reflection: transmitted beam (325 mW) is compressed and focused on the sample, while first (68 mW) and second (57 mW) reflections are used for independent $f_{ceo}$ detection in two common-path $f$-2$f$ interferometers. Both operate at 1.25 µm with frequency doubling in a 2-mm-thick periodically-poled lithium niobate crystal, yielding $f_{ceo}$ beat-note SNR of approximately 75 dB (100 kHz resolution bandwidth).

### CEP stabilization

For the CEP stabilization, the $f_{ceo}$ RF signal (generated either via optically induced currents in the sample, or in the $f$-to-2$f$ interferometer) is low-pass filtered at 1.9 MHz, amplified by 40 dB (DHPVA-101, Femto) and compared by a double-balanced mixer against an RF reference signal from a signal generator (SMA100B, R&S). The error signal from the mixer is fed to a PID servo controlling the pump power via a custom driver of the modulation diode. The residual phase noise curves were measured out-of-loop by an RF spectrum analyzer (FSW, R&S) using the signal from one of the two available $f$-2$f$ interferometers.

### PHz chip samples

We tested two different samples, containing bulk GaN or nanolaminated Ir/$Al_2O_3$ as active media. The former is a commercially available standalone GaN crystal with Fe doping to ensure semi-insulating properties (MSE supplies LLC, WA0220). GaN is advantageous for light-field-driven current generation owing to its high bandgap and high optical nonlinearity. The nanolaminated sample, nano-Ir, is a custom-made one fabricated with atomic-layer-deposition that interleaves nm-thick metallic Ir layers with $Al_2O_3$ (for details see Fehér et al.[33,39]. Both samples have gold electrodes deposited on its surface by means of electron beam lithography. The electrodes form a micrometer-scale gap into which the laser beam is focused. The GaN sample had a 2-µm-wide gap formed by 2-µm-wide electrodes. The Ir sample had a gap with varying widths ranging from 1 to 5 µm following the design presented by Hanus et al.[32]. In both cases the electrodes widen towards contact points through which they were wire-bonded to a printed-circuit board and then through an SMA connector to a preamplifier.

### Electric field-amplitude scan

To obtain power scaling in Fig. 4, the current from the PHz chip was measured with a lock-in amplifier referenced to the $f_{CEO}$ frequency of the phase-locked laser oscillator. The lock-in phase-independent amplitude of the voltage signal $V_0$ was converted to current $J_0$ by considering the gain factor $G$ of the transimpedance preamplifier (DLPCA-200, FEMTO Messtechnik GmbH). Hence, $J_0 = V_0/G$. It was ensured during the measurements that the used $f_{CEO}$ is within the specified bandwidth of the preamplifier.

The on-target field magnitude was varied by a rotation of a polarizer. This effectively attenuates the projection of the field polarization vector on the axis connecting the opposing metallic electrodes of the CEP chip. For every rotation of the polarizer, we performed a small area transverse scan of the sample position and took the maximum signal as a representative value. The amplitude calibration was done by a thermopile-based power meter and an in-situ knife-edge measurement. We obtained 8.2 µm and 5.8 µm for the focal spot size in the horizontal direction for GaN and nano-Ir cases respectively. The various spot sizes are due to the different size of upstream aperture (iris) that regulated the maximum intensity on the target to avoid damage. The knife-edge also allowed us to calibrate the position of the PHz chip of the measurement along the optical axis with respect to the beam focus with micrometer precision. The degree of circularity of the focus was obtained from scanning slit beam-profiler performed before the measurement. See Supplementary Information for details on the beam profiles and sample placement. We then calculate the maximum field assuming Gaussian transversal and temporal profile. The pulse duration is gained from pulse characterization measurements with the frequency-resolved optical gating (FROG) technique yielding 14-fs pulses (intensity full width at half maximum), see Fig.1(c) and Supplementary Information.

**Data availability**

The data supporting the findings of this study are publicly available in the repository cited in ref. [54].

**Acknowledgements**

MK acknowledges the funding from the Max Planck Society within the Max Planck Partner Group Programme.

**Author contributions**

The research was initiated by F.K. and led by P.D. and M.K; M.K. developed and characterized the laser source with contributions from M.P. and J.S.; P.S worked up the sample design; V.H. conceived and led the experiment; V.H. and M.K. performed the experiments and analyzed the data; V.H., M.K. and P.D. wrote the manuscript, which was reviewed by all authors

**Funding**

The project “Ultrastable pulsed lasers covering the spectral range from near to far infrared” (FENG.02.02-IP.05-0069/23) is carried out within the First Team programme of the Foundation for Polish Science co-financed by the European Union under the European Funds for Smart Economy 2021-2027 (FENG). We acknowledge support from the National Research, Development and Innovation Office of Hungary (NKFIH), projects KKP_21-137373 and STARTING_25-152512).

**Competing interests**

The authors declare no competing interests.

**Supplementary information**

The online version contains supplementary material available at XXX.

# Supplementary Information for: Carrier-envelope phase stabilization with a petahertz optical chip

*Václav Hanus[1], Michał Pietrzak[2], Jarosław Sotor[2], Péter Sándor[1],*

*Ferenc Krausz[3,4,5,6], Péter Dombi[1,7,*], Maciej Kowalczyk[2,#]*

[1]HUN-REN Wigner Research Centre for Physics, 1121 Budapest, Hungary
[2]Faculty of Electronics, Photonics and Microsystems, Wrocław University of Science and Technology, 50-370 Wrocław, Poland
[3]Ludwig-Maximilians-Universität München, 80539 München, Germany
[4]Max-Planck-Institut für Quantenoptik, 85748 Garching, Germany
[5]Center for Molecular Fingerprinting, 1093 Budapest, Hungary
[6]The University of Hong Kong, Chong Yuet Ming Physics Building, Pokfulam Road, Hong Kong
[7]Department of Atomic Physics, Budapest University of Technology and Economics, 1111 Budapest, Hungary

*E-mail: dombi.peter@wigner.hu

#E-mail: maciej.kowalczyk@pwr.edu.pl

## Beam focus size determination and sample positioning

The general characterization of the focus was performed with a rotating scanning slit beam profiler (Thorlabs BP209-IR2/M) resulting in horizontal and vertical profile along the beam propagation axis *z* as depicted in Supplementary Fig. 1(a). The resulting beam shows high degree of circularity: $w_0$ = 7 µm in horizontal and vertical direction. The astigmatic walk-off of the horizontal and vertical position is 10 µm, being well below the Rayleigh range of 84 µm.

For correct determination of the on-target intensity during the intensity scans presented in Fig. 4 of the main manuscript, the on-target size of the focus needs to be determined, i.e., the position of the sensitive volume of the PHz chip. To calibrate this, we performed an in-situ knife edge measurement in horizontal direction. The edge of the electrodes used to define the sensitive volume serve as the knife edge. The calibration revealed the mutual positioning of the beam focus and the sample position during the intensity (polarization) scans. The actual focal spot size as influenced by the upstream aperture size is measured as well. Supplementary Fig. 1(b) and Fig. 1(c) shows the measurement for GaN and nano-Ir, respectively.

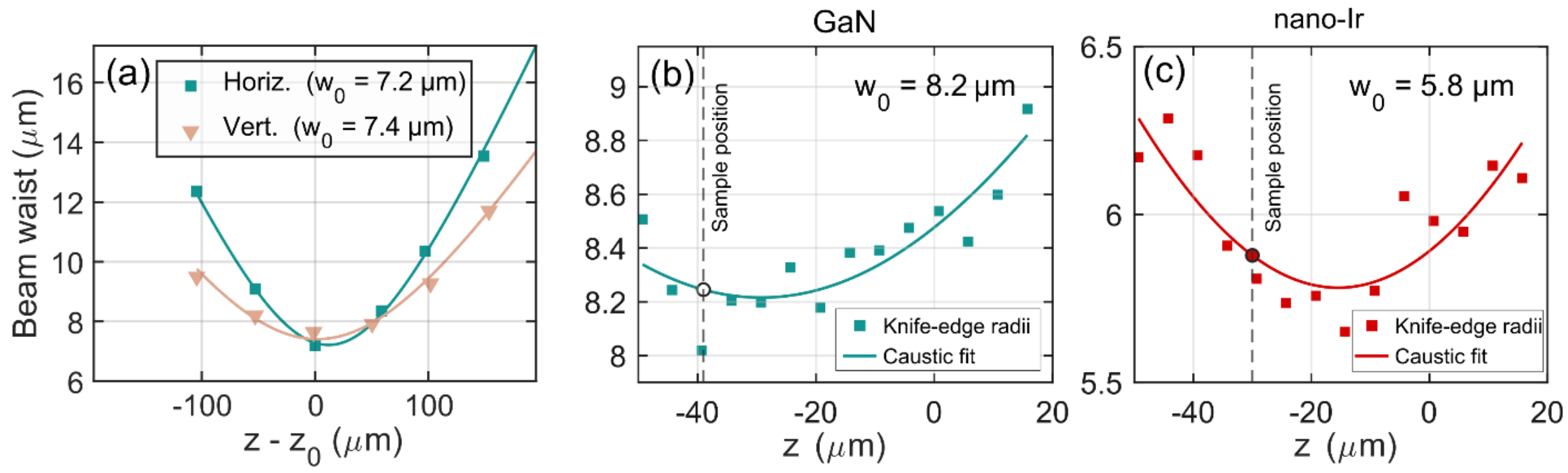

**Supplementary Fig. 1 | Beam size characterization**. **(a)** *Sizes of the beam waist in horizontal and vertical direction as measured with the scanning slit beam profiler.* ***(b,c)*** *In-situ calibration of the sample position using the in-situ knife-edge method as used to calibrate the on-target intensity in Fig.4 main manuscript. Showing the measurements for GaN (b) and nano-Ir (c) samples.*

## Mid-IR pulse characterization

The SHG FROG trace of the optimally compressed pulses at the sample position (after the wedges) and its reconstruction is presented in Supplementary Fig. 2. The FROG error amounts to 0.01.

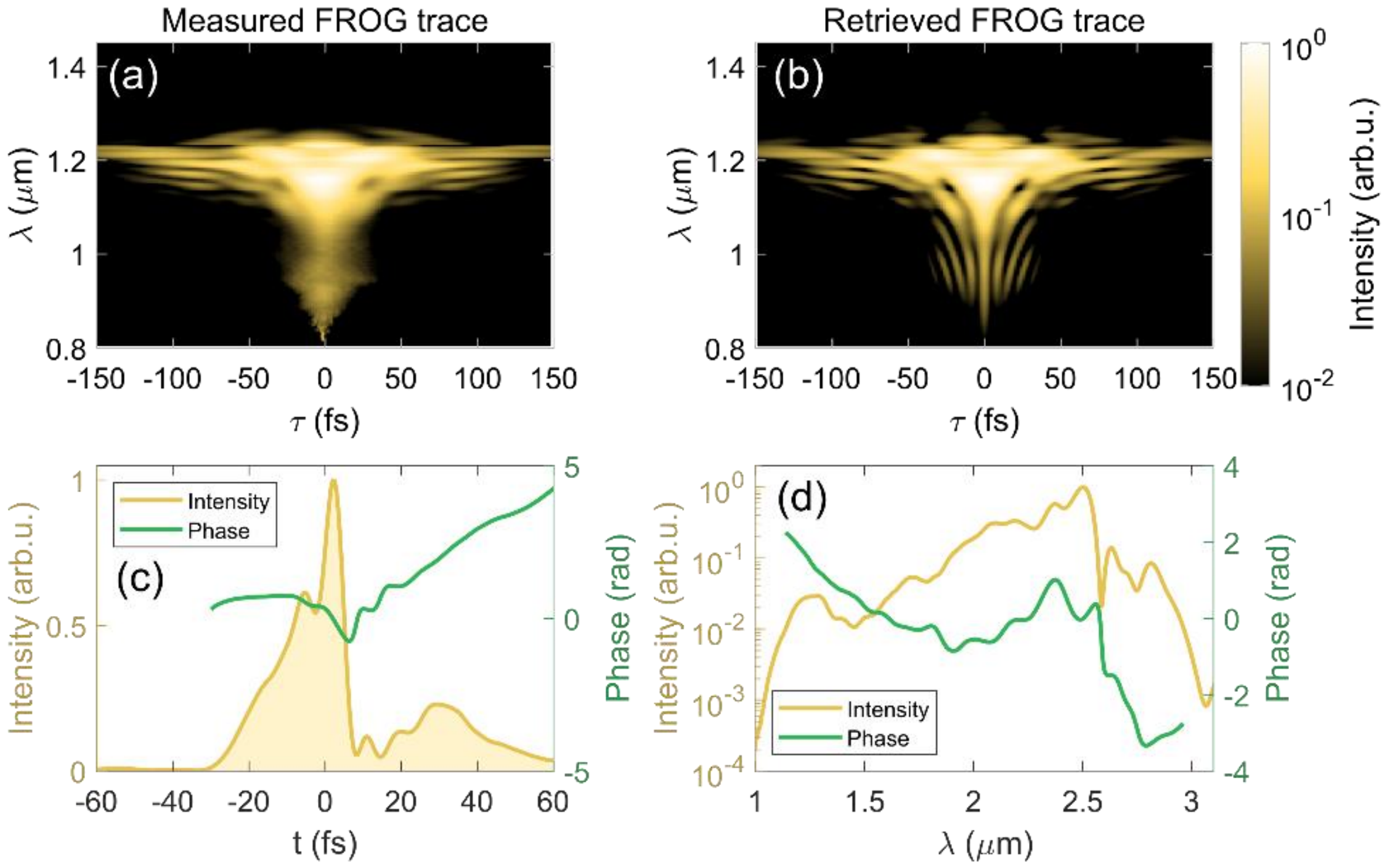

**Supplementary Fig. 2 | SHG FROG pulse characterization.** ***(a)*** *Measured FROG trace and* ***(b)*** *its reconstruction.* ***(c)*** *Pulse intensity profile and temporal phase.* ***(d)*** *Spectral intensity and phase of the reconstructed pulse.*